\documentstyle[11pt]{article}
\def\ft#1#2{{\scriptstyle {#1 \over #2}}}

\def\ww3{{$W_3$}}

\def\del{\partial}

\begin{document}
\topmargin 0pt
\oddsidemargin 5mm
\begin{titlepage}
\begin{flushright}
CTP TAMU-43/94\\
SISSA-143/94/EP\\
hep-th/9410006\\
\end{flushright}
\vspace{1.5truecm}
\begin{center}
{\bf {\Large A Superstring Theory in $2+2$ Dimensions}}
\vspace{1.5truecm}

{\large Z. Khviengia, H. Lu\footnote{Supported in part by the
U.S. Department of Energy, under grant DE-FG05-91-ER40633},\ \  C.N.
Pope$^{1,}$\footnote{Supported in part by the EC Human Capital and Mobility
Programme, under contract number ERBCHBGCT920176.}, E.
Sezgin\footnote{Supported in part by the National Science Foundation, under
grant PHY-9106593}, X.J. Wang and K.W. Xu}
\vspace{1.1truecm}

{\small Center for Theoretical Physics, Texas A\&M University,
                College Station, TX 77843-4242} \vspace{1.1truecm}


\end{center}
\vspace{1.0truecm}

\begin{abstract}
\vspace{1.0truecm}
In this paper we construct a $(2,2)$ dimensional
string theory with manifest $N=1$ spacetime supersymmetry.  We use Berkovits'
approach of augmenting the spacetime supercoordinates by the conjugate momenta
for the fermionic variables.   The worldsheet symmetry algebra is a twisted
and truncated ``small'' $N=4$ superconformal algebra.  The physical spectrum
of the open string contains an infinite number of massless states,
including a supermultiplet of
a self-dual Yang-Mills field and a right-handed spinor and a supermultiplet of
an anti-self-dual Yang-Mills field and a left-handed spinor.  The higher-spin
multiplets are natural generalisations of these self-dual and anti-self-dual
multiplets.
\end{abstract}
\end{titlepage}
\newpage
\pagestyle{plain}
\section{Introduction}

     The Green-Schwarz superstring \cite{GS}, with manifest spacetime
supersymmetry, has  proved to be notoriously difficult to quantise in a
covariant manner.  The difficulty stems from the fact that there is no
kinetic term for the fermionic spacetime coordinates.   This problem has been
overcome recently by Berkovits \cite{Berk} in a reformulation of the
superstring, in which the spacetime supercoordinates are augmented by the
conjugate momenta for the fermionic variables.  The theory has $N=2$
worldsheet supersymmetry, as well as manifest four-dimensional $N=1$ spacetime
supersymmetry.   This theory can be thought of as a ten-dimensional theory
compactified on a Calabi-Yau background.

     It is interesting to investigate whether such an approach could be used
for constructing an intrinsically four-dimensional theory with manifest
spacetime supersymmetry.  This would contrast strikingly with the
$N=2$ NSR string \cite{OV} which, although it has a four-dimensional
spacetime (with $(2,2)$ signature), has no supersymmetry in spacetime.  In
fact it has only one physical state, describing self-dual Yang-Mills in
the open string, and self-dual gravity in the  closed string.  Attempts
have been made to find a supersymmetric version of the theory.  In a recent
paper \cite{BKL}, it was  observed that massless fermionic physical states,
as well as bosonic ones, appear in certain $Z_2$ twisted sectors of the
theory.  There is a kind of twisted $N=2$ spacetime supersymmetry, with two
chiral supercharges whose  anticommutator vanishes.

    If a four-dimensional string of the Berkovits type could be constructed,
it would be quite different from the above case, in that it would have a
manifest spacetime supersymmetry.  A way to build such a theory is suggested
by some work of  Siegel \cite{S1}.  He considered a set  of quadratic
constraints built from the coordinates and momenta of a  superspace in $2+2$
dimensions, and thus displaying manifest spacetime  supersymmetry.  In
\cite{S1} it was proposed that in the case of open  strings, this theory
described self-dual $N=4$ super Yang-Mills, whilst the  corresponding closed
string described self-dual $N=8$ supergravity.

     The full set of constraints considered in \cite{S1} do not generate a
closed algebra.   However, we find that there exists a subset of the
constraints that does close on an algebra, with two bosonic spin-2 generators
and two fermionic spin-2 generators.  In this paper we build a Berkovits-type
open string theory in four dimensions, based on this worldsheet symmetry
algebra.
Noting that the central charge in the ghost sector vanishes, we see that the
matter fields should also have zero central charge.  We achieve this by
taking the coordinates $(X^\mu,\theta^\alpha)$ of a chiral $N=1$ superspace,
together with the canonical momenta $p_\alpha$ for the fermionic
coordinates.  This is a chiral restriction of the analogous matter system
introduced by Berkovits \cite{Berk}.

     The chiral truncation that we are making is possible only if the
signature of the four-dimensional spacetime is $(2,2)$.  In this case, the
$SO(2,2)$ Lorentz group is the direct product $SL(2,R)_{\rm L} \times
SL(2,R)_{\rm R}$, with dotted spinorial indices transforming under
$SL(2,R)_{\rm L}$, and undotted indices transforming under $SL(2,R)_{\rm R}$.
The bosonic currents are singlets under the entire Lorentz group,
but the two fermionic currents form a doublet under $SL(2,R)_{\rm L}$.
This means, as we shall show later, that while the physical states of the
theory will have manifest $SL(2,R)_{\rm R}$ spacetime symmetry, the
$SL(2,R)_{\rm L}$ symmetry, although remaining unbroken, is not always
manifest.

     The spectrum of physical states of this $(2,2)$ dimensional theory
turns out to be quite rich.
Owing to the fact that the fermionic currents, and hence their
associated ghosts, carry $SL(2,R)_{\rm L}$ indices it follows that the
ghost vacuum can transform non-trivially under $SL(2,R)_{\rm L}$.  As a
result, it turns out that the physical spectrum includes an infinite number
of massless states, with arbitrarily high spin.
Included amongst these is a supermultiplet
comprising an anti-self-dual Yang-Mills field and a left-handed spinor.
There is also supermultiplet consisting of a self-dual Yang-Mills field and
a right-handed spinor.  The higher-spin multiplets are natural
generalisations of these anti-self-dual and self-dual multiplets.  Such
higher-spin massless fields seem not to have been encountered previously in
string theory.

\section{The constraint algebra and BRST charge}

     In this section, we set up the algebra of constraints that defines the
string theory, and construct the BRST operator.  The matter system consists
of the four spacetime coordinates $X^{\alpha\dot\alpha} =
\sigma^{\alpha\dot\alpha}_\mu\, X^\mu$, the two-component Majorana-Weyl spinor
$\theta^\alpha$, and its conjugate momentum $p_\alpha$.  In the language of
conformal field theory, these satisfy the OPEs
\begin{equation}
X^{\alpha\dot\alpha}(z) X^{\beta\dot\beta}(w) \sim - \epsilon^{\alpha\beta}
\, \epsilon^{\dot\alpha\dot\beta} \log(z-w),\qquad \qquad p_\alpha(z)
\theta_{\beta}(w) \sim {\epsilon_{\alpha\beta}\over z-w} \ .
\label{matopes}
\end{equation}
When we need to be explicit, we use conventions in which the spacetime metric
is given by
$\eta_{\mu\nu}={\rm diag}\, (-1,-1,1,1)$, the indices $\mu,\nu\,\ldots$ run
from 1 to 4, and the mapping between tensor indices and 2-component spinor
indices is defined by
\begin{equation}
V^{\alpha\dot\alpha}=
\pmatrix{V^{1\dot1} & V^{1\dot2}\cr
         V^{2\dot1} & V^{2\dot2}\cr} = {1\over\sqrt 2}\,
\pmatrix{V^1+V^4 & V^2-V^3 \cr
         V^2+V^3 &-V^1+V^4\cr}\ ,
\label{vdw}
\end{equation}
where $V^\mu$ is an arbitrary vector.  The Van der Waerden symbols
$\sigma^{\alpha\dot\alpha}_\mu$ thus defined satisfy
$\sigma^{\alpha\dot\alpha}_\mu\, \sigma^{\mu\, \beta\dot\beta} =
\epsilon^{\alpha\beta}\, \epsilon^{\dot\alpha\dot\beta}$, where
$\epsilon_{12}=\epsilon^{12}=1$.  Spinor indices are
raised and lowered according to the usual ``North-west/South-east''
convention, with $\psi^\alpha= \epsilon^{\alpha\beta}\, \psi_\beta$ and
$\psi_\alpha=\psi^\beta\, \epsilon_{\beta\alpha}$, {\it etc.}, so we have
$\psi^1=\psi_2$ and $\psi^2=-\psi_1$.  Note that since the indices are
two-dimensional, we have the useful Schoutens identity $X^\alpha\,
Y_\alpha\, Z_\beta + X_\alpha\, Y_\beta\, Z^\alpha + X_\beta\,
Y^\alpha\, Z_\alpha =0$.

     In \cite{S1}, Siegel proposed to build a string theory implementing the
set of constraints given by $\Big\{\del X^{\alpha\dot\alpha}\del
X_{\alpha\dot\alpha},\, p_\alpha\,\del\theta^\alpha, \,p_\alpha p^\alpha,\,
\del\theta_\alpha\, \del\theta^\alpha,\, p_\alpha\,\del
X^{\alpha\dot\alpha},\,
\del\theta_\alpha\, \del X^{\alpha\dot\alpha}\Big\}$.  However, it follows
from (\ref{matopes}) that whilst the second order poles in the OPEs
amongst this set of constraints give back the same set of constraints, not
all the first-order poles can be re-expressed as the derivatives of the
constraints.  In other words, the algebra does not close.  (Note that this
non-closure occurs even at the classical level of Poisson brackets, or
single OPE contractions.)  Accordingly, we
choose a subset of Siegel's constraints that form a closed algebra, namely
\begin{eqnarray}
T &=& -\ft12 \del X^{\alpha\dot\alpha}\, \del X_{\alpha\dot\alpha} - p_\alpha\,
\del\theta^\alpha \ ,\nonumber \\
\label{matcur}
S &=& - p_\alpha\, p^\alpha\ ,\\
G^{\dot\alpha} &=& - p_\alpha\, \del X^{\alpha\dot\alpha}\ .\nonumber
\end{eqnarray}
We see from the energy-momentum tensor that from the worldsheet viewpoint,
$\theta^\alpha$ has conformal weight 0, and $p_\alpha$ has conformal weight
1.  Thus the two bosonic currents $T$ and $S$, and also the two fermionic
currents $G^{\dot\alpha}$, have conformal spin 2.  The currents are all
primary, and the remaining non-trivial OPE is given by
\begin{equation}
G^{\dot\alpha}(z) G^{\dot\beta}(w) \sim {2\epsilon^{\dot\alpha\dot\beta} \,
S\over (z-w)^2} + {\epsilon^{\dot\alpha\dot\beta}\,\del S\over z-w} \ .
\end{equation}
The action for the matter system takes the form
\begin{equation}
I=\int d^2z\, \Big( -\ft12 \del X^{\alpha\dot\alpha}\, \bar\del X_{\alpha
\dot\alpha} + p_\alpha\, \bar\del \theta^\alpha \Big) \ .
\end{equation}

     Turning now to the BRST operator, we begin by introducing the
anticommuting ghosts $(b,c)$ and $(\beta,\gamma)$ for the bosonic currents
$T$ and $S$; and the commuting ghosts $(r^{\dot\alpha}, s_{\dot\alpha})$ for
the fermionic currents $G^{\dot\alpha}$, with $r^{\dot\alpha}(z)
s^{\dot\beta}(w)\sim -\epsilon^{\dot\alpha\dot\beta}\, (z-w)^{-1}$.
All anti-ghosts $(b,\beta,r^{\dot
\alpha})$ have spin 2, and all ghosts $(c,\gamma, s_{\dot\alpha})$ have spin
$-1$.  Straighforward computation leads to the following result for the BRST
operator:
\begin{eqnarray}
Q &=& Q_0 + Q_1 + Q_2\ ,\\
Q_0 &=& \oint c\Big( -\ft12 \del X^{\alpha\dot\alpha}\, \del X_{\alpha\dot
\alpha} - p_\alpha\, \del\theta^\alpha -b\,\del c -2\beta\,\del\gamma
-\del\beta\, \gamma + 2r^{\dot\alpha}\, \del s_{\dot\alpha} + \del r^{\dot
\alpha}\, s_{\dot\alpha}\Big)\ ,\\
Q_1 &=& \ft12 \oint \gamma\, p_\alpha\, p^\alpha \ ,\\
Q_2 &=& \oint \big(s_{\dot\alpha}\, G^{\dot\alpha} -\beta\, s_{\dot\alpha}
\, \del s^{\dot\alpha} \big)\ .
\label{rsbrst}
\end{eqnarray}

(Note that we can augment our
currents (\ref{matcur}) by including $\Big\{\theta_\alpha \theta^\alpha,\
p_\alpha\, \theta^\alpha,\, \theta_\alpha\, \del X^{\alpha\dot\alpha}\Big\}$
as well.  One can easily verify that the resulting currents generate
precisely the ``small'' $N=4$ superconformal algebra, but in a twisted
basis. If the currents are then untwisted, it can easily be seen that the
algebra has central charge $c=6$, and therefore this realisation could not
be used to build a nilpotent $N=4$ BRST operator, which would require
$c=-12$ for criticality.  Thus one cannot build another string theory with
more constraints just by enlarging our set (\ref{matcur}) to give a full
$N=4$ superconformal set of currents.\footnote{ The $N=4$ superconformal
algebra, and its twisted subalgebra generated by (\ref{matcur}), have been
used recently in the construction of a topological $N=4$ string \cite{BV2}.
This is a different kind of string from the one we are discussing, with no
ghosts, and a BRST operator given by a fermionic spin-1 current in the
twisted algebra which was one of the four spin-$\ft32$ currents prior to
twisting.})

     Under $SL(2,R)_{\rm R}$ transformations, corresponding to the
self-dual Lorentz transformations of the undotted indices, the currents are
all invariant.  Thus the generators of $SL(2,R)_{\rm R}$ are simply given
by
\begin{equation}
J^{\alpha\beta} = \oint \Big(-\ft12 X^{(\alpha}{}_{\dot\alpha}\,\del X^{\beta)
\dot\alpha} + p^{(\alpha}\, \theta^{\beta)} \Big) \ ,\\
\end{equation}
and, for an infinitesimal transformation with (symmetric) parameter
$\omega_{\alpha\beta}$, have the action $\delta \psi^\alpha
=[\omega_{\beta\gamma} J^{\beta\gamma}, \psi^\alpha \}=
\omega^\alpha{}_\beta \, \psi^\beta$ on any undotted index.
The $SL(2,R)_{\rm L}$ transformations, on the other hand, which
correspond to anti-self-dual Lorentz rotations of the dotted indices, rotate
the fermionic currents $G^{\dot\alpha}$, and hence the ghosts $(r^{\dot
\alpha},s_{\dot\alpha})$ must rotate also.  It is quite easy to see that the
generators are given by
\begin{equation}
J^{\dot\alpha\dot\beta} = \oint\Big(\ft12 X^{\alpha(\dot\alpha}\,\del X_\alpha
{}^{\dot\beta)} +  r^{(\dot\alpha}\, s^{\dot\beta)}\Big)\ ,\\
\label{asdrot}
\end{equation}
and they transform dotted indices according to $\delta\psi^{\dot\alpha} =
\omega^{\dot\alpha}{}_{\dot\beta}\, \psi^{\dot\beta}$.
These spacetime Lorentz transformations are symmetries of the
two-dimensional action including ghosts, and they commute with the BRST
charge.

    It is also useful to write down the form of the spacetime supersymmetry
generators.  These are given by
\begin{eqnarray}
q_\alpha &=& \oint p_\alpha \ ,\\
q^{\dot\alpha} &=& \oint \Big(- \theta_\alpha\, \del X^{\alpha\dot\alpha} -
\gamma\, r^{\dot\alpha} + b\, s^{\dot\alpha} \Big) \ .
\label{rsssusy}
\end{eqnarray}
The somewhat unusual ghost terms in $q^{\dot\alpha}$ are a consequence of
the fact that $r^{\dot\alpha}$ and $s_{\dot\alpha}$ transform under the
anti-self-dual spacetime Lorentz group.  It is straightforward to verify
using (\ref{matopes}) that these supercharges generate the usual $N=1$
spacetime superalgebra
\begin{equation}
\{q_\alpha,q_\beta\}=0= \{q^{\dot\alpha},q^{\dot\beta} \},\qquad
\{q^\alpha,q^{\dot\alpha} \}= P^{\alpha\dot\alpha} \ ,
\end{equation}
where $P^{\alpha\dot\alpha}=\oint \del X^{\alpha\dot\alpha}$ is the
spacetime translation operator.

     As usual in a theory with fermionic currents, it is appropriate to
bosonise the associated commuting ghosts.  Thus we write
\begin{equation}
r^{\dot\alpha} = \del\xi^{\dot\alpha}\, e^{-\phi_{\dot\alpha}}\ , \qquad
\qquad s_{\dot\alpha}= \eta_{\dot\alpha}\, e^{\phi_{\dot\alpha}}\ ,
\label{boso}
\end{equation}
where $\eta_{\dot\alpha}$ and $\xi^{\dot\alpha}$ are anticommuting fields
with spins 1 and 0 respectively.  The OPEs of the bosonising fields are
$\eta_{\dot\alpha}(z)\xi^{\dot\beta}(w)\sim \delta_{\dot\alpha}^{\dot\beta}
\,(z-w)^{-1}$, and $\phi_{\dot\alpha}(z)\phi_{\dot\beta}(w)\sim
-\delta_{\dot\alpha\dot\beta}\,\log(z-w)$.
Note that the bosonisation breaks the manifest $SL(2,R)_{\rm L}$ covariance,
and that the $\dot\alpha$ index in (\ref{boso}) is not summed.  In view of
this non-covariance, there is no particular advantage in using upper as well
as lower indices on $\phi_{\dot\alpha}$, and we find it more convenient
always to user use lower ones for this purpose.

     The BRST operator is easily re-expressed in terms of the bosonised
fields; the $(r,s)$ terms in $Q_0$ become $\oint c\Big(-\eta_{\dot\alpha}\,
\del \xi^{\dot\alpha} -\ft12(\del\phi_1)^2 -\ft12(\del\phi_2)^2 -\ft32
\del^2\phi_1 -\ft32\del^2\phi_2\Big)$, whilst $Q_2$ becomes
\begin{eqnarray}
Q_2 &=& \oint \Big(\eta_1\, p_\alpha\, \del X^{\alpha 1}\, e^{\phi_1} +
\eta_2\, p_\alpha\, \del X^{\alpha 2}\, e^{\phi_2} \Big)\nonumber \\
&+& \oint\beta\Big(\eta_1\, \del\eta_2 -\del\eta_1\,\eta_2 - \eta_1\, \eta_2
(\del\phi_1 -\del\phi_2)\Big) e^{\phi_1+\phi_2} \ .
\end{eqnarray}
(It is to be understood that an expression such as $e^{\phi_1+\phi_2}$ really
means $:e^{\phi_1}: \, :e^{\phi_2}:$, which equals $-:e^{\phi_2}: \,
:e^{\phi_1}:$ since both of these exponentials are fermions. Thus we have
$e^{\phi_1+\phi_2}=-e^{\phi_2+\phi_1}$ in this rather elliptical notation.)

    The ghost contributions to the $SL(2,R)_{\rm L}$ Lorentz generators
(\ref{asdrot}) become
\begin{eqnarray}
J_+ &=& r_1\, s_1 = \eta_1\, \del\xi^2\, e^{\phi_1-\phi_2}\ ,\nonumber\\
J_- &=& r_2\, s_2 = \eta_2\, \del\xi^1\, e^{-\phi_1+\phi_2}\ ,
\label{asdghost}\\
J_3 &=& r_{(1}\, s_{2)} = -\ft12(\del\phi_1-\del\phi_2)\ .
\end{eqnarray}
One can easily see that these generate an $SL(2,R)$ Kac-Moody algebra.  The
translation of the supersymmetry charges into bosonised form is obtained by
simple substitution.

     Since the zero modes of the $\xi^{\dot\alpha}$ fields are not included
in the Hilbert space of physical states, there exist BRST non-trivial
picture-changing operators $Z^{\dot\alpha}=\{Q,\xi^{\dot\alpha}\}$ which
can give new BRST non-trivial physical operators when normal ordered with
others.  Explicitly, they take the form
\begin{eqnarray}
Z^1 &=& c\, \del\xi^1 - p_\alpha\, \del X^{\alpha 1}\, e^{\phi_1} -
 \Big(2\beta\, \del\eta_2 + \del\beta\,\eta_2 +2\beta\, \eta_2\,\del\phi_2
\Big) e^{\phi_1+\phi_2}\ ,\\
Z^2 &=& c\, \del\xi^2 - p_\alpha\, \del X^{\alpha 2} \, e^{\phi_2} -
 \Big(2\beta\, \del\eta_1 + \del\beta\,\eta_1 +2\beta\, \eta_1\,\del\phi_1
\Big) e^{\phi_1+\phi_2}\ ,
\label{pic}
\end{eqnarray}
Unlike the picture-changing operator in the usual $N=1$ NSR superstring, it
appears that here the operators have no inverse.  This is similar to the
situation in the $N=1$ superstring formulation of \cite{Berk}.

\section{Physical states}

\subsection{Preliminaries}

     In this section, we shall discuss the cohomology of the BRST operator,
and  present some results for physical states in the theory.  Owing to the
rather  unusual feature in this theory that some of the ghosts carry target
spacetime  spinor indices, the notion of the standard ghost vacuum requires
some  modification.  We begin by noting that the non-vanishing
correlation function that defines the meaning of conjugation is given by
\begin{equation}
\Big\langle \del^2c\,\del c\, c\,\del^2\gamma\, \del\gamma\, \gamma\,
e^{-3\phi_1-3\phi_2}\, \theta^2\, \Big\rangle \ne 0\ ,
\label{prod}
\end{equation}
where $\theta^2\equiv\theta^\alpha\, \theta_\alpha$.
In terms of the bosonised form of the commuting
ghosts, the usual operator $e^{-\phi_1-\phi_2}$ appearing in the definition
of the ghost vacuum can be generalised to an operator $W_{\dot\alpha_1\cdots
\dot\alpha_{2s}}$, totally symmetric in its indices, whose component with
$(s+m)$ indices taking the value $\dot1$ and $(s-m)$ taking the value $\dot2$
is given by
\begin{equation}
W_{\dot1\cdots \dot1 \dot2\cdots \dot2} =
\lambda(s,m)\,\del^{s+m-1}\eta_1 \cdots
\del\eta_1 \, \eta_1\,  \del^{s-m-1}\eta_2\cdots
\del\eta_2 \, \eta_2\, e^{(s+m-1)\phi_1 +(s-m-1)\phi_2}\ .
\label{Wdef}
\end{equation}
The normalisation constants $\lambda(s,m)$ are given by
\begin{equation}
\lambda(s,m) = \prod_{p=1}^{s+m-1}\ \prod_{q=1}^{s-m-1} {1\over p!\, q!}\ ,
\end{equation}
where any product over an empty range is defined to be 1. $W$ in (\ref{Wdef})
has $(s+m)$ factors involving $\eta_1$, and $(s-m)$ factors involving
$\eta_2$, with $-s\le m\le s$.  It is the $J_3=m$ state in the
$(2s+1)$-dimensional spin-$s$ representation of $SL(2,R)_{\rm L}$.  The
operator $W_{\dot1\cdots \dot1}$ corresponds to the highest-weight state in the
representation, satisfying $J_+\, W_{\dot1\cdots \dot1}=0$,
with the remaining $2s$
states being obtained by acting repeatedly with $J_-$, each application of
which turns a further ``$\dot1$'' index into a ``$\dot2$'',
until the lowest-weight
state $W_{\dot2\cdots \dot2}$ is obtained.  Note, incidentally,
that the form of the
states given in (\ref{Wdef}) becomes rather simple if one bosonises the
$(\eta,\xi)$ fields.

     We may also define a ``conjugate'' operator $\widetilde
W^{\dot\alpha_1\cdots\dot\alpha_{2s}}$, again totally symmetric in its
indices, by
\begin{equation}
\widetilde W^{\dot1\cdots \dot1 \dot2\cdots \dot2} =
\tilde\lambda(s,m)\,\del^{s+m}\xi^1\cdots
\del^2\xi^1 \, \del\xi^1\,  \del^{s-m}\xi^2\cdots
\del^2\xi^2 \, \del\xi^2 e^{-(s+m+2)\phi_1 -(s-m+2)\phi_2}\ ,
\end{equation}
with
\begin{equation}
\tilde\lambda(s,m)= \prod_{p=1}^{s+m}\ \prod_{q=1}^{s-m} {1\over p!\, q!}\ .
\end{equation}
Thus the usual ghost vacuum operator $e^{-\phi_1-\phi_2}$ and its
``conjugate'' $e^{-2\phi_1-2\phi_2}$ correspond to the $s=0$ cases $W$ and
$\widetilde W$ respectively.  All the operators
$W_{\dot\alpha_1\cdots\dot\alpha_{2s}}$ and $\widetilde
W^{\dot\alpha_1\cdots\dot\alpha_{2s}}$ have worldsheet conformal spin 2, and
they all have the property of defining vacuum states that are annihilated by
the positive Laurent modes of $r^{\dot\alpha}$ and $s_{\dot\alpha}$, but not
by the negative modes.

    These operators have simple properties when acted on by the BRST
operator.  The relevant facts can be summarised in the following lemmas:
\begin{eqnarray}
Q_2\, W_{\dot\alpha_1\cdots \dot\alpha_{2s}}\, \theta^\alpha\,e^{ip\cdot X} &=&
i p^{\alpha\dot\alpha_{2s+1}}\,
W_{\dot\alpha_1\cdots \dot\alpha_{2s+1}}\,
e^{ip\cdot X}\ ,\\
Q_2\, \widetilde W^{\dot\alpha_1\cdots \dot\alpha_{2s}}\, \theta^\alpha\,
e^{ip\cdot X} &=&
i p^{\alpha(\dot\alpha_1}\, \widetilde W^{\dot\alpha_2\cdots
\dot\alpha_{2s})}\,
e^{ip\cdot X}\ .
\label{lemmas}
\end{eqnarray}
A factor of $\gamma$ or $\del\gamma\gamma$ may be included on both sides of
the equation in either formula.  Note that in (\ref{lemmas}), the right-hand
side is defined to be zero if $s=0$.   It is worth remarking that
 we have recovered the manifest
covariance under $SL(2,R)_{\rm L}$ in the expressions for the
$W^{\dot\alpha_1\cdots\dot\alpha_{2s}}$ and $\widetilde W^{\dot\alpha_1
\cdots\dot\alpha_{2s}}$, even though it was broken by the bosonisation of the
ghosts.

\subsection{Supersymmetry transformations}

     The $N=1$ spacetime supersymmetry provides an organising principle for
the physical states.  It is therefore convenient to begin by writing down a
subset of the physical states that can then be filled out into multiplets by
using the supersymmetry generators.  We shall focus here on such a subset
for a class of massless states in the theory.  They are described by the
following operators:
\begin{eqnarray}
U&=&h_{\alpha\dot\alpha_1\cdots \dot\alpha_{2s}}\, c\, \gamma\,
W^{\dot\alpha_1\cdots \dot\alpha_{2s}}\, \theta^\alpha\,
e^{ip\cdot X}\ ,\label{Ustate}\\
\widetilde U&=&\tilde h_{\alpha\dot\alpha_1\cdots \dot\alpha_{2s}}\, c\,
\gamma\,  \widetilde W^{\dot\alpha_1\cdots \dot\alpha_{2s}}\, \theta^\alpha\,
e^{ip\cdot X}\ ,
\label{UUstate}
\end{eqnarray}
For now, we shall derive the supermultiplets associated with these operators
without yet imposing any physical-state conditions.  Having then derived the
corresponding off-shell supersymmetry transformations in this section, we
shall then, in the next section, discuss the physical-state conditions and
the associated on-shell transformation rules.

     Let us first consider the operator $U$ given in (\ref{Ustate}).
Acting with the supersymmetry generators $\epsilon_\alpha\, q^\alpha$ and
$\epsilon_{\dot\alpha}\, q^{\dot \alpha}$, we obtain the complete set of six
operators,
\begin{eqnarray}
U&=&h_{\alpha\dot\alpha_1\cdots \dot\alpha_{2s}}\, c\, \gamma\,
W^{\dot\alpha_1\cdots \dot\alpha_{2s}}\, \theta^\alpha\,
e^{ip\cdot X}\ ,\nonumber\\
V&=&g_{\alpha\dot\alpha_1\cdots \dot\alpha_{2s}}\, c\, \del\gamma\,\gamma\,
W^{\dot\alpha_1\cdots \dot\alpha_{2s}}\, \theta^\alpha\,
e^{ip\cdot X}\ , \nonumber\\
\Psi&=&h_{\dot\alpha_1\cdots \dot\alpha_{2s}}\, c\, \gamma\,
W^{\dot\alpha_1\cdots \dot\alpha_{2s}}\,
e^{ip\cdot X}\ , \nonumber\\
\Phi&=&g_{\dot\alpha_1\cdots \dot\alpha_{2s}}\, c\,\del\gamma\, \gamma\,
W^{\dot\alpha_1\cdots \dot\alpha_{2s}}\,
e^{ip\cdot X}\ , \label{Umult}\\
R&=&b_{\dot\alpha_1\cdots \dot\alpha_{2s}}\, c\, \gamma\,
W^{\dot\alpha_1\cdots \dot\alpha_{2s}}\,\theta^2\,
e^{ip\cdot X}\ , \nonumber\\
S&=&d_{\dot\alpha_1\cdots \dot\alpha_{2s}}\, c\,\del\gamma\, \gamma\,
W^{\dot\alpha_1\cdots \dot\alpha_{2s}}\,\theta^2\,
e^{ip\cdot X}\ ,\nonumber
\end{eqnarray}
The action of the supersymmetry generators can then be written in terms of
this basis of operators.  For example, acting on $U$ with $\epsilon_\alpha\,
q^\alpha$, we find
\begin{equation}
\epsilon_\alpha\,q^\alpha\, U=
\epsilon^\alpha\, h_{\alpha\dot\alpha_1\cdots \dot\alpha_{2s}}\, c\, \gamma\,
W^{\dot\alpha_1\cdots \dot\alpha_{2s}}\,
e^{ip\cdot X}\ ,
\label{examp}
\end{equation}
which can be recognised as an operator of the form $\Psi$.  A similar
pattern can be obtained for the remaining supersymmetry transformations of
the operators.  If we associate spacetime fields with the polarisation
spinors according to the scheme
\begin{eqnarray}
A_{\alpha\dot\alpha_1\cdots \dot\alpha_{2s}}&\leftrightarrow&
h_{\alpha\dot\alpha_1\cdots\dot\alpha_{2s}}\, e^{ip\cdot X}\ ,\qquad
B_{\alpha\dot\alpha_1\cdots \dot\alpha_{2s}}\leftrightarrow
g_{\alpha\dot\alpha_1\cdots\dot\alpha_{2s}}\, e^{ip\cdot X}\ ,\nonumber\\
\psi_{\dot\alpha_1\cdots \dot\alpha_{2s}}&\leftrightarrow&
h_{\dot\alpha_1\cdots\dot\alpha_{2s}}\, e^{ip\cdot X}\ ,\qquad
\phi_{\dot\alpha_1\cdots \dot\alpha_{2s}}\leftrightarrow
g_{\dot\alpha_1\cdots\dot\alpha_{2s}}\, e^{ip\cdot X}\ ,\label{fields}\\
\rho_{\dot\alpha_1\cdots \dot\alpha_{2s}}&\leftrightarrow&
b_{\dot\alpha_1\cdots\dot\alpha_{2s}}\, e^{ip\cdot X}\ ,\qquad
\sigma_{\dot\alpha_1\cdots \dot\alpha_{2s}}\leftrightarrow
d_{\dot\alpha_1\cdots\dot\alpha_{2s}}\, e^{ip\cdot X}\ ,\nonumber
\end{eqnarray}
we can obtain the supersymmetry algebra of the spacetime fields.  For
example, the transformation given in (\ref{examp}) gives rise to
$\delta \psi_{\dot\alpha_1\cdots \dot\alpha_{2s}}= \epsilon^\alpha\,
A_{\alpha\dot\alpha_1\cdots \dot\alpha_{2s}}$.  The complete set of such
supersymmetry transformations takes the form
\begin{eqnarray}
\delta A_{\alpha\dot\alpha_1\cdots \dot\alpha_{2s}} &=&
\epsilon^{\dot\alpha}\, \del_{\alpha\dot\alpha}\,
\psi_{\dot\alpha_1\cdots \dot\alpha_{2s}} + \epsilon_\alpha\,
\rho_{\dot\alpha_1\cdots \dot\alpha_{2s}}\ ,\nonumber\\
\delta\psi_{\dot\alpha_1\cdots \dot\alpha_{2s}}&=&
\epsilon^\alpha\, A_{\alpha\dot\alpha_1\cdots \dot\alpha_{2s}}\ ,
\nonumber\\
\delta \rho_{\dot\alpha_1\cdots \dot\alpha_{2s}}&=&
\epsilon_{\dot\alpha}\, \del^{\alpha\dot\alpha}\,
A_{\alpha\dot\alpha_1\cdots \dot\alpha_{2s}}\ ,\nonumber\\
\delta B_{\alpha\dot\alpha_1\cdots \dot\alpha_{2s-1}}&=&
-\epsilon^{\dot\alpha_{2s}}\, A_{\alpha\dot\alpha_1\cdots \dot\alpha_{2s}}
+ \epsilon^{\dot\alpha}\, \del_{\alpha\dot\alpha}\,
\phi_{\dot\alpha_1\cdots \dot\alpha_{2s-1}} + \epsilon_\alpha\,
\sigma_{\dot\alpha_1\cdots \dot\alpha_{2s-1}}\ ,\label{stsusy}\\
\delta\phi_{\dot\alpha_1\cdots \dot\alpha_{2s-1}}&=&
\epsilon^{\dot\alpha_{2s}}\, \psi_{\dot\alpha_1\cdots \dot\alpha_{2s}} +
\epsilon^\alpha\, B_{\alpha\dot\alpha_1\cdots \dot\alpha_{2s-1}}\ ,
\nonumber\\
\delta\sigma_{\dot\alpha_1\cdots \dot\alpha_{2s-1}}&=&
\epsilon^{\dot\alpha_{2s}}\, \rho_{\dot\alpha_1\cdots \dot\alpha_{2s}} +
\epsilon_{\dot\alpha}\, \del^{\alpha\dot\alpha}\,
B_{\alpha\dot\alpha_1\cdots \dot\alpha_{2s-1}}\ .\nonumber
\end{eqnarray}
One can easily verify that these transformations close off-shell on the usual
$N=1$ supersymmetry algebra.  In the next subsection, we shall show how this
algebra truncates to an on-shell algebra of the physical fields, when we
impose the physical-state conditions.

  We now discuss the supermultiplet associated with the operator $\widetilde
U$ given in (\ref{UUstate}).  In an analogous manner to the above, we obtain
five additional operators by acting with the supersymmetry generators. Thus
in all we have
\begin{eqnarray}
\widetilde U&=&\tilde h_{\alpha\dot\alpha_1\cdots \dot\alpha_{2s}}\, c\,
\gamma\,  \widetilde W^{\dot\alpha_1\cdots \dot\alpha_{2s}}\, \theta^\alpha\,
e^{ip\cdot X}\ ,\nonumber\\
\widetilde V&=&\tilde g_{\alpha\dot\alpha_1\cdots \dot\alpha_{2s}}\, c\,
\del\gamma\,\gamma\, \widetilde W^{\dot\alpha_1\cdots \dot\alpha_{2s}}\,
\theta^\alpha\,  e^{ip\cdot X}\ , \nonumber\\
\widetilde\Psi&=&\tilde h_{\dot\alpha_1\cdots \dot\alpha_{2s}}\, c\, \gamma\,
\widetilde W^{\dot\alpha_1\cdots \dot\alpha_{2s}}\,\theta^2\,
e^{ip\cdot X}\ , \nonumber\\
\widetilde\Phi&=&\tilde g_{\dot\alpha_1\cdots \dot\alpha_{2s}}\,
c\,\del\gamma\, \gamma\,  \widetilde W^{\dot\alpha_1\cdots
\dot\alpha_{2s}}\,\theta^2\,   e^{ip\cdot X}\ , \label{UUmult}\\
\widetilde R&=&\tilde b_{\dot\alpha_1\cdots \dot\alpha_{2s}}\, c\, \gamma\,
\widetilde W^{\dot\alpha_1\cdots \dot\alpha_{2s}}\,
e^{ip\cdot X}\ , \nonumber\\
\widetilde S&=&\tilde d_{\dot\alpha_1\cdots \dot\alpha_{2s}}\, c\,
\del\gamma\,
\gamma\, \widetilde  W^{\dot\alpha_1\cdots \dot\alpha_{2s}}\,
e^{ip\cdot X}\ ,\nonumber
\end{eqnarray}
We associate spacetime fields with the polarisation spinors precisely as in
(\ref{fields}), except that now all quantities carry tildes.  The
supersymmetry transformation rules are given by
\begin{eqnarray}
\delta\widetilde A_{\alpha\dot\alpha_1\cdots \dot\alpha_{2s}} &=&
\epsilon^{\dot\alpha}\, \del_{\alpha\dot\alpha}\,
\tilde\rho_{\dot\alpha_1\cdots \dot\alpha_{2s}} + \epsilon_\alpha\,
\widetilde\psi_{\dot\alpha_1\cdots \dot\alpha_{2s}}\ ,\nonumber\\
\delta\widetilde\psi_{\dot\alpha_1\cdots \dot\alpha_{2s}}&=&
\epsilon_{\dot\alpha}\,\del^{\alpha\dot\alpha}\, \widetilde
A_{\alpha\dot\alpha_1\cdots \dot\alpha_{2s}}\ , \nonumber\\
\delta \tilde\rho_{\dot\alpha_1\cdots \dot\alpha_{2s}}&=&
\epsilon^\alpha\,
\widetilde A_{\alpha\dot\alpha_1\cdots \dot\alpha_{2s}}\ ,\nonumber\\
\delta \widetilde B^\alpha{}_{\dot\alpha_1\cdots \dot\alpha_{2s+1}}&=&-
\epsilon_{(\dot\alpha_{2s+1}}\, \widetilde A^\alpha{}_{\dot\alpha_1\cdots
\dot\alpha_{2s})} - \epsilon_{\dot\alpha}\, \del^{\alpha\dot\alpha}\,
\tilde\sigma_{\dot\alpha_1\cdots \dot\alpha_{2s+1}} + \epsilon^\alpha\,
\tilde \phi_{\dot\alpha_1\cdots \dot\alpha_{2s+1}}\ ,\label{stsuusy}\\
\delta\tilde\phi_{\dot\alpha_1\cdots \dot\alpha_{2s+1}}&=&
\epsilon_{(\dot\alpha_{2s+1}}\, \widetilde\psi_{\dot\alpha_1\cdots
\dot\alpha_{2s})} + \epsilon^{\dot\alpha}\,
\del_{\alpha\dot\alpha}\, \widetilde B^\alpha{}_{\dot\alpha_1\cdots
\dot\alpha_{2s+1}}\ , \nonumber\\
\delta\tilde \sigma_{\dot\alpha_1\cdots
\dot\alpha_{2s+1}}&=& \epsilon_{(\dot\alpha_{2s+1}}\,
\tilde\rho_{\dot\alpha_1\cdots \dot\alpha_{2s})} - \epsilon_{\alpha}\,
\widetilde B^\alpha{}_{\dot\alpha_1\cdots \dot\alpha_{2s+1}}\ .
\nonumber
\end{eqnarray}

\subsection{Physical States}

     Having established the off-shell structure of the supermultiplets for
the operators $U$ and $\widetilde U$ in (\ref{Ustate}, \ref{UUstate}), we
turn to an analysis of the physical-state conditions for these operators.
There are two parts to this analysis; first requiring that the operators be
annihilated by the BRST operator, and then investigating the conditions
under which they are BRST non-trivial.

     Beginning with the untilded operators, we find that $U$ itself is
annihilated by $Q$ provided that the following conditions hold:
\begin{equation}
p^{\alpha\dot\alpha}\, p_{\alpha\dot\alpha}=0\ ,\quad \qquad
h^{\alpha(\dot\alpha_1\cdots\dot\alpha_{2s}}\, p_\alpha{}^{\dot\alpha)} =0\ .
\label{eqh}
\end{equation}
The first of these is just the mass-shell condition for massless states.
Having ensured that $U$ is annihilated by $Q$, we must also check to see
whether it is BRST non-trivial.  One way to do this is by constructing
conjugate operators that have a non-vanishing inner product with $U$, as
defined by (\ref{prod}).  If the inner-product is non-vanishing for
conjugate operators that are annihilated by $Q$, then $U$ is BRST
non-trivial.  Operators $U^\dagger$ conjugate to $U$ have the form
\begin{equation}
U^\dagger = f_{\alpha\dot\alpha_1\cdots\dot\alpha_{2s}}\, \del c\, c\,
\del\gamma\, \gamma\, \widetilde W^{\dot\alpha_1\cdots\dot\alpha_{2s}}\,
\theta^\alpha\ , e^{ip\cdot X}\ ,
\end{equation}
which is annihilated by $Q$ if
\begin{equation}
p^{\alpha\dot\alpha_1}\, f_{\dot\alpha_1\cdots\dot\alpha_{2s}}=0\ .
\label{eqf}
\end{equation}
It is convenient to choose a particular momentum frame in order to analyse
the true physical degrees of freedom that are implied by these kinematical
conditions.  The null momentum vector $p^\mu$ may, without loss of
generality, be chosen to be $p^\mu=(1,0,0,1)$.  From (\ref{vdw}), this
implies that all components of $p^{\alpha\dot\alpha}$ are zero except for
$p^{1\dot1}=\sqrt2$.  In this frame, the solutions to (\ref{eqh}) and
(\ref{eqf}) are
\begin{equation}
h_{1\dot\alpha_1\cdots\dot\alpha_{2s}}=0\ ,\qquad
f_{1 \dot1\dot\alpha_2\cdots\dot\alpha_{2s}} =0 \ .
\label{solhf}
\end{equation}
The inner product has the form $\langle U^\dagger\, U\rangle= f^{\alpha
\dot\alpha_1\cdots\dot\alpha_{2s}}\, h_{\alpha \dot\alpha_1\cdots\dot
\alpha_{2s}}=
f^{2\dot1\cdots \dot1}\, h_{2\dot1\cdots \dot1}$.  Thus there is just one
physical degree of freedom described by $U$, corresponding to the
polarisation spinor component $h_{2\dot1\cdots \dot1}$.  The other
non-vanishing components of $h_{\alpha\dot\alpha_1\cdots\dot\alpha_{2s}}$
allowed by
(\ref{solhf}) correspond to BRST trivial states, and can be expressed back in
covariant language as pure-gauge states with
\begin{equation}
h_{\alpha\dot\alpha_1\cdots\dot\alpha_{2s}}=p_{\alpha(\dot\alpha_1}\,
\Lambda_{\dot\alpha_2\cdots\dot\alpha_{2s})}\ ,\label{hguage}
\end{equation}
where $\Lambda_{\dot\alpha_2\cdots\dot\alpha_{2s}}$ is arbitrary.
We note also, for
future reference, that the equation of motion for
$h^{\alpha\dot\alpha_1\cdots\dot\alpha_{2s}}$ in (\ref{eqh}) is equivalent to
\begin{equation}
h^{\alpha\dot\alpha_1\cdots\dot\alpha_{2s}}\, p_\alpha{}^{\dot\alpha} =0\ .
\label{nonsym}
\end{equation}

     The operator $\Psi$ in (\ref{Umult}) is annihilated by $Q$ provided
just that the mass-shell condition $p^{\alpha\dot\alpha}\,
p_{\alpha\dot\alpha}=0$ is satisfied.  To see the physical degrees of
freedom, we again consider conjugate operators $\Psi^\dagger$, which have
the form $\Psi^\dagger=f_{\dot\alpha_1\cdots\dot\alpha_{2s}}\,
\del c\, c\, \del
\gamma\, \gamma\, \widetilde W^{\dot\alpha_1\cdots\dot\alpha_{2s}}\,
\theta^2\,
e^{ip\cdot X}$.  This is annihilated by $Q$ provided that $p^{\alpha\dot
\alpha_1}\, f_{\dot\alpha_1\cdots\dot\alpha_{2s}}=0$. In the special momentum
frame, the solution is $f_{\dot1\dot\alpha_2\cdots\dot\alpha_{2s}}=0$. Thus the
inner product is proportional to $h^{\dot2\cdots \dot 2}\,
f_{\dot2\cdots \dot 2}$, so only the one component  $h^{\dot2\cdots \dot 2}$
describes a true physical degree of freedom.  The unphysical BRST-trivial
components correspond to polarisation spinors of the pure-gauge form
\begin{equation}
h^{\dot\alpha_1\cdots\dot\alpha_{2s}}=p^{\alpha(\dot\alpha_1}\,
\Lambda_\alpha{}^{\dot\alpha_2\cdots\dot\alpha_{2s})}\ .\label{gguage}
\end{equation}

     The analysis of the operators $V$ and $\Phi$ in (\ref{Umult}) is
precisely the same as the above analyses for $U$ and $\Psi$ respectively.
The operators $R$ and $S$ in (\ref{Umult}) are annihilated by $Q$ only if
$b_{\dot\alpha_1\cdots\dot\alpha_{2s}}=0$ and $d_{\dot\alpha_1\cdots\dot
\alpha_{2s}}=0$, and thus they do not describe any physical degrees of
freedom.  The corresponding spacetime fields
$\rho_{\dot\alpha_1\cdots\dot\alpha_{2s}}$ and
$\sigma_{\dot\alpha_1\cdots\dot\alpha_{2s}}$ can be thought of as auxiliary
fields that are needed in order to have a supersymmetry algebra
(\ref{stsusy}) that closes off-shell.  In fact one can easily see in the
off-shell transformations (\ref{stsusy}) that it is consistent to set
$\rho_{\dot\alpha_1\cdots\dot\alpha_{2s}}$ and $\sigma_{\dot\alpha_1\cdots
\dot\alpha_{2s}}$
to zero, since the right-hand sides of their variations then vanish by virtue
of the field equations for $A_{\alpha\dot\alpha_1\cdots\dot\alpha_{2s}}$ and
$B_{\alpha\dot\alpha_1\cdots\dot\alpha_{2s-1}}$, which can be read off from
(\ref{nonsym}).  The on-shell supersymmetry transformations are thus given
by the remaining four variations in (\ref{stsusy}), with
$\rho_{\dot\alpha_1\cdots\dot\alpha_{2s}}$ and $\sigma_{\dot\alpha_1\cdots
\dot\alpha_{2s}}$ set to zero, namely
\begin{eqnarray}
\delta A_{\alpha\dot\alpha_1\cdots \dot\alpha_{2s}} &=&
\epsilon^{\dot\alpha}\, \del_{\alpha\dot\alpha}\,
\psi_{\dot\alpha_1\cdots \dot\alpha_{2s}} \ ,\nonumber\\
\delta\psi_{\dot\alpha_1\cdots \dot\alpha_{2s}}&=&
\epsilon^\alpha\, A_{\alpha\dot\alpha_1\cdots \dot\alpha_{2s}}\ ,
\nonumber\\
\delta B_{\alpha\dot\alpha_1\cdots \dot\alpha_{2s-1}}&=&
-\epsilon^{\dot\alpha_{2s}}\, A_{\alpha\dot\alpha_1\cdots \dot\alpha_{2s}}
+ \epsilon^{\dot\alpha}\, \del_{\alpha\dot\alpha}\,
\phi_{\dot\alpha_1\cdots \dot\alpha_{2s-1}}  \ ,\label{stsusysh}\\
\delta\phi_{\dot\alpha_1\cdots \dot\alpha_{2s-1}}&=&
\epsilon^{\dot\alpha_{2s}}\, \psi_{\dot\alpha_1\cdots \dot\alpha_{2s}} +
\epsilon^\alpha\, B_{\alpha\dot\alpha_1\cdots \dot\alpha_{2s-1}}\ .
\nonumber
\end{eqnarray}
In the special momentum frame that we
discussed above, the physical degrees of freedom are described by the
components $A_{2\dot1\dot1\cdots\dot1}$, $B_{2\dot1\cdots\dot1}$,
$\psi_{\dot1\dot1\cdots\dot1}$ and $\phi_{\dot1\cdots\dot1}$ of the
spacetime fields.

    It is instructive to examine the supermultiplet in more detail in the
special case $s=\ft12$.  The on-shell supersymmetry transformations now
take the form
\begin{eqnarray}
\delta A_{\alpha\dot\alpha} &=& \epsilon^{\dot\beta}\, \del_{\alpha
\dot\beta}\, \psi_{\dot\alpha} \ ,\nonumber\\
\delta\psi_{\dot\alpha} &=& \epsilon^\alpha\, A_{\alpha\dot\alpha}\ ,
\nonumber\\
\delta B_\alpha &=& -\epsilon^{\dot\alpha}\, A_{\alpha\dot\alpha} +
\epsilon^{\dot\alpha}\, \del_{\alpha\dot\alpha}\, \phi\ ,\label{s12}\\
\delta\phi &=& \epsilon^{\dot\alpha}\, \psi_{\dot\alpha} + \epsilon^\alpha \,
B_\alpha\ .\nonumber
\end{eqnarray}
The multiplet is reducible.  First, we note that $\{A_{\alpha\dot\alpha},
\psi_{\dot\alpha}\}$ satisfy a closed algebra of spins
$\{(\ft12,\ft12),(0,\ft12)\}$.  If instead we set $A_{\alpha\dot\alpha}$ and
$\psi_{\dot\alpha}$ to zero, we get an irreducible multiplet
$\{B_\alpha,\phi\}$ of spins $\{(\ft12,0),(0,0)\}$.  A third irreducible
multiplet can be obtained by setting instead $B_\alpha=0$, which implies that
$A_{\alpha\dot\alpha}=\del_{\alpha\dot\alpha}\phi$.  This corresponds to a
multiplet $\{\psi_{\dot\alpha},\phi\}$ with spins $\{(0,\ft12),(0,0)\}$.
A fourth irreducible multiplet can be obtained by defining $\widehat
A_{\alpha \dot\alpha} = A_{\alpha\dot\alpha} - \del_{\alpha\dot\alpha}\,
\phi$.  This corresponds to a multiplet $\{\widehat A_{\alpha\dot\alpha},
B_\alpha\}$ of spins $\{(\ft12,\ft12), (\ft12,0)\}$.
     This reducibility of the supermultiplet occurs for all values of $s$,
giving rise to analogous irreducible multiplets of spins
$\{(\ft12,s),(0,s)\}$, $\{(\ft12,s-\ft12),(0,s-\ft12)\}$,
$\{(0,s),(0,s-\ft12)\}$ and $\{(\ft12,s),(\ft12, s-\ft12)\}$ respectively.

    The field equation for $A_{\alpha\dot\alpha}$ in the $s=\ft12$ case,
which follows from the physical-state condition (\ref{eqh}), is
$\del^{\alpha(\dot\alpha}\, A_\alpha{}^{\dot\beta)}=0$.  This is invariant
under gauge transformation $A_{\alpha\dot\alpha}\rightarrow
A_{\alpha\dot\alpha} + \del_{\alpha\dot\alpha} \, \Lambda$.  Thus the
gauge-invariant field strength $F_{\mu\nu}=\del_\mu A_\nu-\del_\nu A_\mu=
F_{\alpha\beta}\epsilon_{\dot\alpha\dot\beta} +F_{\dot\alpha\dot\beta}
\epsilon_{\alpha\beta}$ has $F_{\dot\alpha\dot\beta}\equiv
\del_{\alpha(\dot\alpha} A^\alpha{}_{\dot\beta)}=0$, whilst
$F_{\alpha\beta}\equiv -\del_{(\alpha}{}^{\dot\alpha} A_{\beta)\dot\alpha}$
is non-zero.  It therefore corresponds to a self-dual gauge field.  Note
that the superpartner $\psi_{\dot\alpha}$ is left-handed, and consequently
they form a non-standard realisation of the supersymmetry algebra, as given
in (\ref{s12}).  The unusual dimension for the spinor field
$\psi_{\dot\alpha}$ is related to the fact that it does not satisfy the usual
Dirac equation.  However, we can define a new spinor field $\chi_\alpha =
\del_{\alpha\dot\alpha}\, \psi^{\dot\alpha}$, which does satisfy the Dirac
equation.  In the special momentum frame, the physical degree of freedom is
carried by the component $\chi_2$.  In terms of $\chi_\alpha$, the
supersymmetry transformations for the irreducible multiplet can be written in
the standard form
\begin{eqnarray}
\delta A_{\alpha\dot\alpha} &=& \epsilon_{\dot\alpha}\, \chi_\alpha\ ,
\nonumber \\
\delta \chi_\alpha &=& \epsilon^\beta\, F_{\alpha\beta}
\ .
\end{eqnarray}
In obtaining this transformation rule from (\ref{s12}), we have dropped pure
gauge terms, and made use of the physical-state conditions.

     Similar redefinitions can be performed for higher values of $s$.  In
this case, we define $C_{\alpha_1\cdots\alpha_{2s}\dot\alpha}=
\del_{(\alpha_2}{}^{\dot\alpha_2}\cdots \del_{\alpha_{2s})}{}^{\dot
\alpha_{2s}}\, A_{\alpha_1\dot\alpha_2\cdots\dot\alpha_{2s}\dot\alpha}$, and
$\chi_{\alpha_1\cdots\alpha_{2s}}=
\del_{\alpha_1}{}^{\dot\alpha_1}\cdots \del_{\alpha_{2s}}{}^{\dot
\alpha_{2s}}\,\psi_{\dot\alpha_1\cdots\dot\alpha_{2s}}$.  The transformation
rules become, after dropping pure-gauge terms and using the physical-state
conditions,
\begin{eqnarray}
\delta C_{\alpha_1\cdots\alpha_{2s}\dot\alpha} &=&
\epsilon_{\dot\alpha}\, \chi_{\alpha_1\cdots\alpha_{2s}}\ ,\nonumber\\
\delta \chi_{\alpha_1\cdots\alpha_{2s}} &=&
\epsilon^{\alpha_{2s+1}}\, F_{\alpha_1\cdots\alpha_{2s+1}}\ ,\label{redsusy}
\end{eqnarray}
where $F_{\alpha_1\cdots\alpha_{2s+1}}\equiv
\del_{(\alpha_1}{}^{\dot\alpha}\, C_{\alpha_2\cdots\alpha_{2s+1})\dot\alpha}$
is a generalised self-dual field strength.  It is worth remarking that even
though the redefinitions involve derivatives of the original fields, the
on-shell degrees of freedom are preserved, as can be easily verified by
using the special momentum frame.

      At this point, a comment is in order.   The supersymmetry transformation
rules given in (\ref{stsusysh}) and (\ref{redsusy}) close on-shell for all
values of $s$.   This implies that sometimes a boson is transformed into
the derivative of a fermion.  However, the unusual dimensions of the fields
implied by this are not inconsistent;  they reflect the fact that one
member of the supermultiplet does not satisfy a conventional equation of
motion.
The original field $A_{\alpha\dot\alpha_1\cdots\dot\alpha_{2s}}$ satisfies the
field equation implied by (\ref{eqh}),
\begin{equation}
\del_{\alpha(\dot\alpha}A^\alpha{}_{\dot\alpha_1\cdots\dot\alpha_{2s})} = 0\ .
\end{equation}
The original field $\psi_{\dot\alpha_1\cdots\dot\alpha_{2s}}$ does not have
a covariant first order field equation.   These original fields have
gauge invariances implied by (\ref{hguage}) and (\ref{gguage}),
\begin{equation}
\delta A_{\alpha\dot\alpha_1\cdots\alpha_{2s}} = \del_{\alpha(\dot
\alpha_1}\, \Lambda_{\dot\alpha_2\cdots\dot\alpha_{2s})}\ ,\qquad
\delta \psi_{\dot\alpha_1\cdots\dot\alpha_{2s}} =\del_{\alpha(\dot\alpha_1}\,
\Lambda^\alpha{}_{\dot\alpha_2\cdots\dot\alpha_{2s})}\ .
\end{equation}
The situation is reversed for the redefined fields.  The field $C_{\alpha_1
\cdots\alpha_{2s}\dot\alpha}$ does not have a covariant first order
field equation whilst the field $\chi_{\alpha_1\cdots\alpha_{2s}}$ satisfies
the Dirac-type equation
\begin{equation}
\del^{\alpha_1\dot\alpha}\, \chi_{\alpha_1\cdots\alpha_{2s}} =0\ .
\end{equation}
There is no gauge symmetry for the $\chi_{\alpha_1\cdots\alpha_{2s}}$ field.
In fact the equation of motion implies that there is only one on-shell degree
of freedom.  In the special momentum frame, it is $\chi_{2\cdots 2}$.   The
field $C_{\alpha_1\cdots\alpha_{2s}\dot\alpha}$ has the gauge invariance
\begin{equation}
\delta C_{\alpha_1\cdots\alpha_{2s}\dot\alpha} =
\del_{(\alpha_1|\dot\alpha|}\, \lambda_{\alpha_2\cdots\alpha_{2s})}\ .
\end{equation}
The situation for the $B_{\alpha\dot\alpha_1\cdots\dot\alpha_{2s-1}}$ and
$\phi_{\dot\alpha_1\cdots\dot\alpha_{2s-1}}$ fields is completely analogous.

   We now turn to the tilded physical operators of the forms given by
(\ref{UUmult}).  The operators $\widetilde R$ and $\widetilde S$ are
annihilated by $Q$ provided simply that the mass-shell condition is
satisfied.  However, these operators are BRST trivial, which can be seen by
showing that there exist no conjugate operators that are annihilated by $Q$.
The operator $\widetilde V$ is annihilated by $Q$ provided that the
mass-shell condition is satisfied and that the polarisation spinor satisfies
\begin{equation}
p^{\alpha\dot\alpha_1}\, \tilde g_{\alpha\dot\alpha_1\cdots\dot\alpha_{2s}}
=0\ .
\end{equation}
Similarly, the polarisation spinor in the operator $\widetilde \Phi$ must
satisfy
\begin{equation}
p^{\alpha\dot\alpha_1}\, \tilde g_{\dot\alpha_1\cdots\dot\alpha_{2s}}=0\ .
\end{equation}
In the special momentum frame introduced earlier, this condition implies
that only the component $\tilde g_{\dot2\cdots\dot2}$ is non-zero.  By
considering operators conjugate to $\widetilde V$, it is easy to see that
only the component $\tilde g_{1\dot2\cdots\dot2}$ in $\widetilde V$ describes
a BRST non-trivial physical state.

     The physical-state condition for $\widetilde U$ is similar to that for
$\widetilde V$, and the only BRST non-trivial physical state is
given by the $\tilde h_{1\dot2\cdots\dot2}$ component.  However, the
analysis for $\widetilde\Psi$ is more complicated.  By itself, $\widetilde
\Psi$ can never be annihilated by $Q$, but together with an additional term,
$\widetilde\Psi$ can become physical.  The form of the new operator is
\begin{equation}
\widehat\Psi = \widetilde\Psi +
\tilde t_{\alpha\dot\alpha_1\cdots\dot \alpha_{2s+1}}\, c\,
\del\gamma\,\gamma\, \widetilde  W^{\dot\alpha_1\cdots\dot\alpha_{2s+1}}\,
\theta^\alpha\, e^{ip\cdot X} \ .
\end{equation}
Note that the additional term is of precisely the same off-shell form as
$\widetilde V$, so introducing this term does not affect the discussion of
the supersymmetry transformation rules (\ref{stsuusy}).  On shell, however,
the polarisation  spinor $\tilde t_{\alpha\dot\alpha_1\cdots
\dot\alpha_{2s+1}}$ does not satisfy the same condition as $\tilde
g_{\alpha\dot\alpha_1 \cdots\dot \alpha_{2s+1}}$.  The physical-state
condition for $\widehat\Psi$ implies
\begin{eqnarray}
p^{\alpha\dot\alpha_1}\, \tilde h_{\dot\alpha_1\cdots\dot\alpha_{2s}} &=&0
\ ,\nonumber\\
\tilde h_{\dot\alpha_1\cdots\dot\alpha_{2s}}&=& p^{\alpha\dot\alpha_{2s+1}}\,
\tilde t_{\alpha\dot\alpha_1\cdots\dot\alpha_{2s+1}}\ .
\label{newstate}
\end{eqnarray}
In the special momentum frame, this new  BRST non-trivial state is described
by the component $\tilde h_{\dot2\cdots\dot2}= \sqrt2 \,
\tilde t_{1\dot1\dot2\cdots\dot2}$. Thus the complete on-shell multiplet of
tilded fields comprises $\widetilde B_{1\dot2\dot2\cdots\dot2}$,
$\tilde\phi_{\dot2\dot2\cdots\dot2}$, $\widetilde A_{1\dot2\cdots \dot2}$
and $\widetilde\psi_{\dot2\cdots\dot2}$.  The supersymmetry
transformation rules for these fields can be read off from (\ref{stsuusy}).
Reverting to covariant notation, they take the form
\begin{eqnarray}
\delta \widetilde A_{\alpha\dot\alpha_1\cdots\dot\alpha_{2s}} &=&
\epsilon_\alpha\, \widetilde \psi_{\dot\alpha_1\cdots\dot\alpha_{2s}}\ ,
\nonumber\\
\delta \widetilde \psi_{\dot\alpha_1\cdots\dot\alpha_{2s}}&=&
\epsilon_{\dot\alpha}\,\del^{\alpha\dot\alpha}\,
\widetilde A_{\alpha\dot\alpha_1\cdots\dot\alpha_{2s}}\,\nonumber\\
\delta \widetilde B^\alpha{}_{\dot\alpha_1\cdots\dot\alpha_{2s+1}}&=&
-\epsilon_{(\dot\alpha_{2s+1}}\,
\widetilde A^\alpha{}_{\dot\alpha_1\cdots\dot\alpha_{2s})}+
\epsilon^\alpha\, \tilde\phi_{\dot\alpha_1\cdots\dot\alpha_{2s+1}}\ ,
\nonumber\\
\delta \tilde\phi_{\dot\alpha_1\cdots\dot\alpha_{2s+1}} &=&
\epsilon_{(\dot\alpha_{2s+1}}\,
\widetilde\psi_{\dot\alpha_1\cdots\dot\alpha_{2s})}
+ \epsilon^{\dot\alpha}\, \del_{\alpha\dot\alpha}\,
\widetilde B^\alpha{}_{\dot\alpha_1\cdots\dot\alpha_{2s+1}}\ .
\end{eqnarray}
The situation here is different from that for the untilded on-shell
transformations in that pure-gauge terms have been introduced by re-writing
the on-shell transformations in a covariant form.    Consequently, the
on-shell algebra closes only modulo pure-gauge terms, in addition to the
use of the field equations.    The reason for this difference is that in
the untilded case, the physical-state conditions implied that the $R$ and
$S$ operators given in (\ref{Umult}) were zero.  In the tilded case, on
the other hand, the physical-state conditions imply instead that the
$\widetilde R$ and $\widetilde S$ operators are BRST trivial.  Thus they
describe gauge degrees of freedom.   The $\widetilde \psi_{\dot\alpha_1
\cdots\dot\alpha_{2s}}$ field satisfies a Dirac-type field equation:
\begin{equation}
\del^{\alpha\dot\alpha_1}\, \widetilde\psi_{\dot\alpha_1
\cdots\dot\alpha_{2s}}=0\ .
\end{equation}
For the same reason that has been explained for the untilded case, the gauge
field $\widetilde A_{\alpha\dot\alpha_1\cdots\dot\alpha_{2s}}$ does not
satisfy a covariant field equation.  However, the physical-state conditions
imply that
\begin{eqnarray}
\del^{\alpha\dot\alpha_1}\,
\widetilde A_{\alpha\dot\alpha_1\cdots\dot\alpha_{2s}}
 &=&0\ ,\nonumber\\
\delta \widetilde A_{\alpha\dot\alpha_1\cdots\dot\alpha_{2s}} &=&
\del_{\alpha}{}^{\dot\alpha_{2s+1}}\,
\widetilde \Lambda_{\dot\alpha_1\cdots\dot\alpha_{2s+1}} \ .
\end{eqnarray}
The discussion for the fields $\tilde \phi_{\dot\alpha_1
\cdots\dot\alpha_{2s}}$ and $\widetilde B_{\alpha\dot\alpha_1\cdots
\dot\alpha_{2s}}$ is precisely the same as the above.

     The on-shell supersymmetry algebra is reducible.  Firstly, we note that
$\{\widetilde A_{\alpha\dot\alpha_1\cdots \dot\alpha_{2s}},
\widetilde \psi_{\dot\alpha_1\cdots \dot\alpha_{2s}}\}$ form
an irreducible supermultiplet with spins $\{(\ft12,s),(0,s)\}$.
If instead we set these two fields to zero, then the fields
$\{\widetilde B_{\alpha\dot\alpha_1\cdots \dot\alpha_{2s+1}},
\tilde\phi_{\dot\alpha_1\cdots \dot\alpha_{2s+1}}\}$ form an irreducible
multiplet with spins $\{(\ft12,s+\ft12),(0,s+\ft12)\}$.
In the special case $s=\ft12$, the former multiplet describes an
anti-self-dual vector gauge field and a left-handed spinor.  Their
supersymmetry transformation rules are
\begin{eqnarray}
\delta \widetilde A_{\alpha\dot\alpha} &=& \epsilon_\alpha\,
\widetilde\psi_{\dot\alpha} \ ,\nonumber\\
\delta \widetilde\psi_{\dot\alpha} &=& \epsilon^{\dot\beta}\,
\widetilde F_{\dot\alpha\dot \beta}\ ,
\end{eqnarray}
where $\widetilde
F_{\dot\alpha\dot\beta}=\del_{\alpha(\dot\alpha}\,\widetilde A^\alpha
{}_{\dot\beta)}$ is the anti-self-dual field strength.  This multiplet
generalises for higher values of $s$, with a generalised anti-self-dual field
strength $\widetilde F_{\dot\alpha_1\cdots\dot\alpha_{2s+1}}=
\del_{\alpha(\dot\alpha_1}\, \widetilde A^\alpha{}_{\dot\alpha_2\cdots
\dot\alpha_{2s+1})}$.   A similar discussion applies to the other multiplet
$\{\widetilde B_{\alpha\dot\alpha_1\cdots\dot\alpha_{2s+1}},
\tilde\phi_{\dot\alpha_1\cdots\dot\alpha_{2s+1}}\}$.

    So far, we have concentrated on massless states in the physical
spectrum.  There are also massive physical states, an example
being $c\, e^{-\phi_1-\phi_2}\, e^{ip\cdot X}$ with $p^{\alpha\dot\alpha}\,
p_{\alpha\dot\alpha}=-2$, implying (mass)$^2 = 2$.  Further examples are
\begin{equation}
V= c\, \del^{2n}\beta\cdots \del\beta\, \beta\, (\del^n p)^2\cdots (\del
p)^2\, p^2\, e^{n(\phi_1+\phi_2)} \, e^{ip\cdot X}\ ,
\end{equation}
where $p^2=p^\alpha\, p_\alpha$, {\it etc}.  These spacetime scalar states
are physical for arbitrary integer $n$, with (mass)$^2 = 2(n+1)(2n+3)$.

      Another class of physical states in the theory is associated with
infinite-dimensional representations of $SL(2,R)_{\rm L}$. Consider, for
example, the operator
\begin{equation}
V=c\, \del\gamma\, \gamma\, \theta^2\,e^{-2\phi_1-\phi_2}\, e^{ip\cdot X}\ .
\label{VB21}
\end{equation}
This is annihilated by the BRST operator provided that the mass-shell
condition $p^{\alpha\dot\alpha} p_{\alpha\dot\alpha}=0$ is satisfied,
together with the transversality condition $p^{\alpha \dot2}=0$.  This
condition is not covariant with respect to $SL(2,R)_{\rm L}$, suggesting
that further terms should be added in order to construct a
fully-covariant physical operator.  This is analogous to viewing a physical
operator built using $W_{\dot\alpha_1\cdots\dot\alpha_{2s}}$ as consisting
of the term involving $W_{\dot1\cdots \dot1}$ plus the remaining $2s$ terms
obtained by acting repeatedly on this highest-weight state with $J_-$.
Thus, noting that $e^{-2\phi_1-\phi_2}$ is a highest-weight state, $J_+\,
e^{-2\phi_1-\phi_2} =0$, we may replace (\ref{VB21}) by the $SL(2,R)_{\rm
L}$ covariant operator
\begin{equation}
V=\sum_{n\ge 0} h_n\, c\, \del\gamma\,\gamma\, \theta^2\, \Big((J_-)^n
e^{-2\phi_1-\phi_2}\Big)\, e^{ip\cdot X}\ .
\end{equation}
One can easily see from the form of the generator $J_-$ in (\ref{asdghost})
that in this case the process of repeated application of $J_-$ will never
terminate, and the sum over $n$ will be an infinite one, corresponding to an
infinite-dimensional representation of $SL(2,R)_{\rm L}$.  The
physical-state conditions will now give a transversality condition on the
components $h_n$ of the polarisation tensor, rather than the non-covariant
condition $p^{\alpha\dot 2}=0$ that resulted when only the $n=0$ term was
included.  The occurrence of infinite-dimensional representations of
$SL(2,R)_{\rm L}$ seems to be an undesirable feature of the theory, and one
may hope that some sort of a truncation may be possible in which such
physical states are projected out of the spectrum.  An understanding of this
point presumably will depend upon knowing the detailed form of the
interactions in the theory.

\section{Discussion}

     In this paper, we have constructed a superstring theory in
four-dimensional spacetime with $(2,2)$ signature, using the Berkovits'
approach of augmenting the spacetime supercoordinates by the conjugate momenta
for the fermionic variables \cite{Berk}.  The form of the theory,
and its local worldsheet symmetries, was motivated by Siegel's proposal
\cite{S1} for a set of constraints that could give rise to self-dual super
Yang-Mills theory or supergravity in $2+2$ dimensions.  Such a theory might
provide one way to generalise the results of Ooguri and Vafa \cite{OV}, who
showed that by starting with an NSR-type string with $N=2$ worldsheet
supersymmetry, one obtains a theory whose physical spectrum describes
(purely bosonic) self-dual Yang-Mills or gravity.  In the theory that we
have considered, $N=1$ spacetime supersymmetry is manifest in the
formulation, as is the right-handed $SL(2,R)$ factor of the $SO(2,2)\equiv
SL(2,R)_{\rm L} \times SL(2,R)_{\rm R}$ Lorentz group.  Owing to the fact
that the fermionic constraint carries an $SL(2,R)_{\rm L}$ spacetime spinor
index, the $SL(2,R)_{\rm L}$ symmetry of the physical spectrum, although
present, is not always manifest.   It is however manifest for the
massless states of arbitrary spin that we considered in this paper.

     The constraints that we have used are a subset of Siegel's constraints
\cite{S1} that form a closed algebra under commutation.  They give rise
to a string theory with an infinite number of massless
states with arbitrary spin.  This
feature can be attributed to the fact that the fermionic constraint carries
an $SL(2,R)_{\rm L}$ spinor index, leading to the existence of ghost vacua
with arbitrary spin under $SL(2,R)_{\rm L}$.

     A full analysis of the physical spectrum will require an understanding
of the interactions of the physical states.  There are two conditions to build
a non-vanishing $n$-point function.  One is the conservation law of momentum
of the physical operators.  The other is that the product of the physical
operators has to include the structure that gives a non-vanishing
inner product, as given in (\ref{prod}). Interactions
amongst the physical states that we have found so far are not easy to come
by.  One example that can occur is a three-point interaction between two
fermions and a boson in a scalar supermultiplet corresponding to the $s=0$
operators $\langle U U \Psi \rangle$, as given in (\ref{Umult}).  For higher
values of $s$, interactions necessarily involve tilded physical states and
picture-changing operators.  However all the tilded physical states have
vanishing nomal-order product with picture-changing operators, and thus we
expect that there are no interactions among these states.   It is of
interest to have a full analysis of the physical spectrum and interactions
of the theory.

\bigskip\bigskip
\centerline{\bf Acknowledgements}
\bigskip

    H.L.\ and C.N.P.\ are grateful to SISSA, Trieste, and E.S. is grateful
to the ICTP, Trieste, for hospitality during the course of this work.

\vfill\eject

\end{document}